\documentclass[amsmath,twocolumn,amssymb,showpacs,showkeys,floatfix,prl]{revtex4-1}

\usepackage{graphicx}
\usepackage{verbatim}
\usepackage{mathtools}
\usepackage{url}
\usepackage{amsthm} 

\usepackage{hyperref} 

\usepackage{cleveref} 

\usepackage{subfigure}

\usepackage{xfrac}
\usepackage{nicefrac}
\newcommand{\KL}[2]{D({#1}\|{#2})}

\usepackage[final]{changes} 
\definechangesauthor[name={Nathaniel Rupprecht}, color=darkgray]{ncr}
\setremarkmarkup{(#2)}
 
\begin{document}

\title{Limits on Inferring the Past}

\author{Nathaniel Rupprecht, Dervis C. Vural}
\email{Corresponding Author: dvural@nd.edu}
\affiliation{University of Notre Dame}

\date{\today}

\begin{abstract}
Here we define and study the properties of retrodictive inference. We derive equations relating retrodiction entropy and thermodynamic entropy, and as a special case, show that under equilibrium conditions, the two are identical. We demonstrate relations involving the KL-divergence and retrodiction probability, and bound the time rate of change of retrodiction entropy. As a specific case, we invert various Langevin processes, inferring the initial condition of \(N\) particles given their final positions at some later time. We evaluate the retrodiction entropy for Langevin dynamics exactly for special cases, and find that one's ability to infer the initial state of a system can exhibit two possible qualitative behaviors depending on the potential energy landscape, either decreasing indefinitely, or asymptotically approaching a fixed value. We also study how well we can retrodict points that evolve based on the logistic map. We find singular changes in the retrodictivity near bifurcations. Counterintuitively, the transition to chaos is accompanied by maximal retrodictability.
\end{abstract}

\maketitle

\section{Introduction}
Many astonishing facts about the origin of the universe, evolution of life, or history of civilizations will never be directly observed, but will only be inferred in the light of their manifestations in the present. Evolved forward in time, any state of knowledge, regardless of how exact, will invariably deteriorate into an entropy maximizing probability distribution
\cite{jaynes1957informationA,jaynes1957informationB, shannon1998mathematical,leff2014maxwell}. How rapidly does our knowledge of the past, as inferred from a measurement made in the present, deteriorate, going backwards in time?

While methods exist for inferring the origin of an observed final state
\cite{box2011bayesian,welling2011bayesian,desmarais2012statistical,nguyen2017morphological}, or inferring some original data after it has been corrupted \cite{hansen2006deblurring,chan2010multilevel} we know little about how accurately the initial state of a many-body system can be characterized given its present state, how quickly a system forgets its initial state due to thermal fluctuations, and how the limit our ability to infer the past depends on system parameters. The answers to these questions should lie in non-equilibrium statistical mechanics, where thermal motion is incorporated into mechanical laws \cite{ullersma1966exactly,yu2015composite,coffey2012langevin}. In systems where thermal collisions erase the information pertaining past states of particles, Fokker-Planck equation constitutes the groundwork of nonequilibrium analysis \cite{wolf1988lie,hashemi2015group,bernstein1984supersymmetry,carrillo1998exponential,toscani1999entropy,schwammle2007general,plastino1997minimum}.

\added{Here we determine the theoretical limits to inferring the initial state of a system, to which we refer as ``retrodiction'' -- in contrast to prediction}. We quantify the quality of retrodiction in terms of retrodiction entropy, $S_R$. We derive a relationship between thermodynamic entropy and retrodiction entropy, and report a lower bound on its generation rate. \added{Then, to apply these ideas to a specific problem,} we consider a collection of particles coupled to a thermal bath, and obtain the time dependence of $S_R$ in convex, concave and flat potentials. \added{To establish whether chaos fundamentally influences retrodictability, we also investigate the retrodiction entropy of the logistic map as it transitions from the non-chaotic regime to the chaotic regime. Finally, we conclude our discussion with a comparison of retrodiction entropy to other inverse statistical methods and methods for comparing predictability and retrodictability.}

\section{Definitions and Notation}
\added{Our system consists of a set states \(\Omega\), a prior distribution on the set of states, \(P_0\), and a ``transition probability'' function \(\mathcal{T}\). The state space \(\Omega\) will depend on the problem at hand, it could for example be the space of all possible positions and velocities of a collection of particles (i.e. phase space). The prior distribution specifies how the system will be initialized - \(P_0(\alpha)\) is the probability that the system will be prepared in the state \(\alpha \in \Omega\). The transition probability \(\mathcal{T}(\omega \vert \alpha ; t) \) is the probability that the system ends in the state \(\omega \in \Omega \) given that it started in the state \(\alpha \in \Omega\) and evolved for a time \(t\). We will generally suppress the time variable.}

The probability \( \mathcal{R}(\alpha|\omega ; t) = \mathcal{R}_\omega(\alpha) \) that the initial state was $\alpha$ given the final state $\omega$, is given by the Bayes theorem,
\begin{align}
\hspace{-0.1in}\mathcal{R}(\alpha \vert \omega ; t) = \frac{\mathcal{T}(\omega \vert \alpha ; t) P_0(\alpha)}{P_t(\omega)} = \frac{\mathcal{T}(\omega \vert \alpha ; t) P_0(\alpha)}{\sum_{\alpha'} \mathcal{T}(\omega \vert \alpha' ; t) P_0(\alpha^\prime)}.
\label{RetrodictionEntropy}
\end{align}
where \(P_t\) is the prior distribution $P_0$ evolved forwards in time. \added{\(\mathcal{R}\) would typically be called the likelihood or the posterior distribution. In the present context, we will refer to it as the retrodiction probability, and define the entropy associated with it as the retrodiction entropy,}
\begin{align}
S_R(\omega) = - \sum_{\alpha} \mathcal{R}_\omega(\alpha) \log \mathcal{R}_\omega(\alpha).
\label{ent}
\end{align}
Intuitively, the larger $S_R(\omega)$ is, the less accurately the initial state can be inferred given a measurement of the final state, \(\omega\). 

Note that $S_R$ is a function of the final state observed after a single realization of a stochastic process. If the process were to be run again, the particles would end up elsewhere, and have a different $S_R$ associated with that final state. As such, it will be useful to define $S_R$ averaged over all possible final measurements, $\langle S_R\rangle$.

A related quantity of interest is the Kullback-Leibler (KL) divergence \(\KL{p}{q} = \sum_x p(x) \log p(x)/q(x) \), measures the amount of overlap between two distributions $p(x)$ and $q(x)$ \cite{cover2012elements}. Thus, another useful measure of retrodictability is the KL divergence \(\KL{\mathcal{R}_\omega}{P_0}\) between \(\mathcal{R}\) and \(P_0\), which quantifies the amount of information gained over the prior upon a measurement. As our ability to infer the past decreases, the retrodiction probability coincides more with the prior probability, the KL divergence decreases. Ultimately, $\KL{\mathcal{R}_\omega}{P} = 0$ as the measurement \(\omega\) provides no additional information regarding the initial state beyond what we already know; the prior, $P$.

\subsection{\added{Notation}} 
Throughout, we denote the average over all free parameters by \( \langle \cdot \rangle\). \added{However, there are two different types of averages that are indicated by this notation: averages over the distribution on initial states, and averages over distribution on final states. When we average over quantities where the free variable ranges over initial states, we use a probability weight \(P_0\) for each such free variable. For quantities where the free variable ranges over final states, we use a probability weight \(P_t\) for each such free variable. In the case where there are multiple states that are being averaged over, we include a subscript to indicate that there is a free variable to be averaged over.} For example,
\begin{align*}
\langle S_R \rangle &= \sum_{\omega} P_t(\omega) S_R(\omega)
\\
\langle S_T \rangle &= \sum_{\alpha} P_0(\alpha) S_T(\alpha)
\\
\langle \KL{\mathcal{T}_{\alpha_1}}{\mathcal{T}_{\alpha_2}} \rangle &= \sum_{\alpha_1, \alpha_2} P_0(\alpha_1) P_0(\alpha_2) \KL{\mathcal{T}_{\alpha_1}}{\mathcal{T}_{\alpha_2}}
\end{align*}

\section{General Properties of Retrodiction}

\subsection{Relation between retrodiction and thermodynamics}
\added{To facilitate readability onwards, we expose only the crucial steps in the main text, leaving the proofs and derivations to the appendices.} 

Our first key result is the relationship between retrodiction entropy and thermodynamic entropy
\begin{align}
\langle S_R \rangle = \langle S_T \rangle - (S_t - S_0).
\label{FundLemma}
\end{align}
Here $\langle S_T \rangle$ is the average entropy associated with the transition probability $\mathcal{T}_\alpha(\omega)$, whereas $S_0$ and $S_t$ are the entropies associated with the prior probability $P_0$, and the observation probability, $P_t$. Eq. (\ref{FundLemma}) relates our ability to infer the past, $\langle S_R\rangle$, to our ability to predict the future, $\langle S_T \rangle$ and $S_t$. This identity is derived in Appendix A.

Note that Eq. (\ref{FundLemma}) holds for processes both in or out of equilibrium, and provides useful insights on the general properties of $S_R$. For short times, $P_t\simeq P_0$, so $\lim_{t \rightarrow 0+} \langle S_R \rangle / \langle S_T \rangle = 1$. For long times, if the system converges to a stationary distribution $P_\infty$, (as is the case in a bounded space or trapping potential), then $P_t$ and $\mathcal{T}_\alpha(\omega)$ must approach $P_\infty$ independent of the starting state, and (\ref{FundLemma}) implies $\lim_{t\to\infty}\langle S_R \rangle=S_0$, i.e. we cannot guess the initial state any better than using  whatever we already knew before making the measurement.

As another interesting special case, we consider what happens if the prior probability \(P_0\) coincides with the stationary state probability $P_\infty$ (assuming one exists). Then $S_t = S_0$ for all times \(t\), and (\ref{FundLemma}) implies
\begin{equation}
\langle S_R(t) \rangle = \langle S_T(t) \rangle.
\label{EquilibriumSr}
\end{equation}
For example, if we are inferring the past of a system in equilibrium we would be drawing the initial state of the system out of the equilibrium distribution, i.e. using $P_0(s) = e^{-\beta E(s)}/Z$ as the prior probability, measure the positions of some particles, and ask where they used to be. Eq. (\ref{EquilibriumSr}) tells us that in equilibrium, the rate of thermodynamic entropy and retrodiction entropy generation is the same. Our ability to predict the future fades at exactly the same rate as our ability to infer the original state of the system. 

No such correspondence need hold for non-equilibrium processes. For a system with equilibrium entropy $S_{\mathrm{eq}}$, if \(S_0 > S_\mathrm{eq}\) then \(S_t\) will decrease from \(S_0\) at \(t=0\) to \(S_\mathrm{eq}\) as \(t \rightarrow \infty\). Thus \(\langle S_R \rangle > \langle S_T \rangle\). In this case, we know that particles will gather, so we know better where they will be in the future than where they were originally.  In contrast, if \(S_0 < S_\mathrm{eq}\), \(S_t\) will increase in time and \(\langle S_R \rangle < \langle S_T \rangle\). Here, we know more about where the particles were originally than where they will be in the future. To sum up, the more certain we can be about the state of the system in the future, the less certain we are about where the system started out in the past.

\subsection{Experimental measurement of retrodictability}
\added{
It is instructive to view (\ref{FundLemma}) from a practical, empirical perspective. Consider a system of particles evolving in a potential energy landscape $U(\vec{x})$ while coupled to a heat bath. Can we estimate bounds on \(\langle S_R \rangle\) without knowing the microscopic dynamics of the system (e.g. the interparticle interactions) or the potential energy landscape, but only using thermodynamic measurements?}

\added{This is possible under certain conditions. We can initialize a system such that particles are in state \(\alpha\) with probability $P_0(\alpha)$, let the particles evolve for a time \(t\), calorimetrically obtain the change in thermodynamic entropy via $\Delta S_\alpha=\int_\alpha dQ/T$, and then average this over multiple instances to obtain \(\langle S_T \rangle_s\) (the sample average of entropy). The identity \(dS = dQ/T\) holds when the system moves along a reversible path. While it is not trivial to measure \(S_t\) for processes out-of-equilibrium, we can use the equilibrium result, $\langle S_T \rangle = \langle S_R \rangle$ (eq. \ref{EquilibriumSr}) and the second law, to place an upper bound on average retrodiction entropy, for any process (in or out of equilibrium),
\[
\langle S_R\rangle <\langle S_T \rangle_s + S_0.
\]
}
\added{
Under special conditions, we can do better than an inequality. If the prior distribution is uncorrelated \(P_0(x_1,\ldots,x_N) = p(x_1) p(x_2) \ldots p(x_N)\), and if interactions between particles are negligible, then 
\[
P_t(y_1,\ldots,y_N) = \prod_{k=1}^N \left( \sum_{x_k} \mathcal{T}(y_k \vert x_k ; t) p(x_k) \right) \equiv \prod_k q(y_k)
\]
Since each term in this product is independent, the entropy is extensive \(S_t = N H[q]\), and \(S_0 = N H[p]\). Thus, an experimentalist can measure \(S_t - S_0\) by placing \(M\gg1\) particles with a number density \(p(x)\), allow the particles to evolve for a time \(t\), and again calorimetrically integrate $\Delta S=\int dQ/T$ to obtain \(S_t - S_0 \simeq N \Delta S/M\). Note that since \(M\gg1\), \(\Delta S\) will be deterministic. Thus from (\ref{FundLemma}) the retrodictability becomes a difference of two entropy measurements,
\begin{equation}
\langle S_R \rangle = \langle\Delta S \rangle_s - \frac{N}{M} \Delta S,
\label{MeasuringSR}
\end{equation}
The first term on the right is measured by initializing particles individually at $\alpha$ with probability $P_0(\alpha)$ and averaging all outcomes, whereas the second term, by a single shot measurement of a gas initialized with density \(P_0(\alpha)\). We emphasize that this experimental protocol to obtain (\ref{MeasuringSR}) will be valid only when inter-particle interactions are negligible, and for an uncorrelated prior, but as long as these assumptions hold, \(\langle S_R \rangle \) can be known by only performing thermodynamic measurements, without needing to know the underlying potential or microscopic dynamics.}

\subsection{Continuous space and divergence relations} 
For a continuous state space, we may consider $S_R$ to be a differential entropy, which is not invariant under a change of variables. In contrast, $\KL{\mathcal{R}_\omega}{P}$ is invariant under changes of variables, and therefore may be a more desirable measure. We derive, in a similar manner to (\ref{FundLemma}),
\begin{align*} 
\langle \KL{\mathcal{R_\xi}}{P} \rangle = S_t - \langle S_T \rangle = S_0 - \langle S_R \rangle.
\end{align*}
Markovian stochastic processes are known to have a KL-divergence that are non-increasing in time \cite{cover2012elements}. Thus we are motivated to ask how the KL divergence between two forward processes \(\mathcal{T}_\alpha\), compares to the KL divergence between two retrodiction probabilities \(\mathcal{R}_\omega\). First, we show (cf. Appendix \added{B})
\begin{align*}
\langle \KL{\mathcal{R}_{\omega_1}}{\mathcal{R}_{\omega_2}} \rangle &= \langle \KL{P_t}{\mathcal{T}_{\alpha}} \rangle + S_t - \langle S_T \rangle 
\\
\langle \KL{P_t}{\mathcal{T}_{\alpha}} \rangle &= \langle \KL{\mathcal{T}_{\alpha_1}}{\mathcal{T}_{\alpha_2}} \rangle  + \langle S_T \rangle - S_t.
\end{align*}
Combining these gives us the relationship
\begin{align}
\langle \KL{\mathcal{R}_{\omega_1}}{\mathcal{R}_{\omega_2}} \rangle = \langle \KL{\mathcal{T}_{\alpha_1}}{\mathcal{T}_{\alpha_2}} \rangle. 
\label{DklRvsT}
\end{align}
Thus, the average amount of overlap between different retrodiction probability distributions is exactly equal to the average amount of overlap between different forward distributions (cf. Appendix B). Taking the time derivative of both sides tells us that the average rate of increase is the same for forward and reverse probabilities, and that this quantity is non-increasing \cite{cover2012elements}. \added{In Appendix B, we list all the KL divergence relations between the distributions \(\mathcal{T}\), \(\mathcal{R}\), \(P_0\), and \(P_t\).}
\subsection{Lower bound to retrodiction entropy generation} 
We can establish a lower bound on the time rate of change of retrodiction entropy in terms of forward entropies and KL divergences. Differentiating (\ref{FundLemma}) and using the convexity of \(\log\) gives us an upper bound on the rate of change of ${S}_t$ (cf. Appendix C),
\begin{align}
&\dot{S}_t \le \langle \dot{S}_T \rangle + \frac{\partial}{\partial t} \langle \KL{\mathcal{T}_{\alpha_1}}{\mathcal{T}_{\alpha_2}} \rangle - \langle \frac{\partial}{\partial t} \KL{P_0}{\mathcal{R_\omega}} \rangle.
\label{SNTimeInequ}
\end{align}

Using the theorem on Markov processes, we know that the second term in (\ref{SNTimeInequ}) is $\leq0$. The last term in (\ref{SNTimeInequ}) measures the divergence between the prior state and the retrodiction probability, which should decrease with time as the reconstructed probability approaches the prior. Rearranging (\ref{SNTimeInequ}), we get,
\begin{align*}
\frac{\partial}{\partial t} \langle S_R \rangle &\ge -\frac{\partial}{\partial t} \langle \KL{\mathcal{T}_{\alpha_1}}{\mathcal{T}_{\alpha_2}} \rangle + \langle \frac{\partial}{\partial t} \KL{P}{\mathcal{R_\omega}} \rangle
\end{align*}

\subsection{\added{Information theoretical interpretation}}
\added{
From an information theoretic point of view, retrodiction entropy is the amount of information required to specify which state the system was initialized, given an observation of its final state. The KL divergence between the retrodiction probability \(\mathcal{R}_\omega\), and the prior distribution \(P_0\) is a measure of how much information has been gained by making a measurement (above and beyond the information contained in the prior). The KL divergence is asymmetric in its arguments, \(\KL{\mathcal{R}_\omega}{P_0} \ne \KL{P_0}{\mathcal{R}_\omega} \). However, there is a good reason for preferring $\KL{\mathcal{R}_\omega}{P_0}$ over $\KL{P_0}{\mathcal{R}_\omega}$. Letting \(X_0\), \(X_t\) be the random variables for the configuration at times \(0\) and \(t\), it can be shown that
\(\langle \KL{\mathcal{R}_\omega}{P_0} \rangle = I(X_0 ; X_t)\) where \(I(\cdot,\cdot)\) is the mutual information.
In other words, the average KL divergence between retrodiction probabilities and the prior is the mutual information between the initial and final states of the system. We can use this and our other formulas to write retrodiction entropy in terms of mutual information,
\begin{equation}
\langle S_R \rangle = S_0 - I(X_0 ; X_t) = H(X_0) - I(X_0 ; X_t)
\label{MutualInfo}
\end{equation}
}
\added{
While it is impossible to evaluate quantities like \(\KL{\mathcal{R}_\omega}{P_0}\) or \(S_R(\omega)\) for a specific \(\omega\) without being given a specific problem (and being able to evaluate the transition probabilities for that problem), \cref{RetrodictionEntropy,ent,FundLemma,EquilibriumSr,DklRvsT,SNTimeInequ,MutualInfo} hold true quite generally, for any system in or out of equilibrium.
}

\section{Retrodiction of Brownian Particles in a Potential} 
Following these general results, we now study a specific physical system, the retrodiction entropy of Brownian particles diffusing in a potential. The $\alpha^\mathrm{th}$ coordinate (\(\alpha\) = x, y, z, ...) of the \(k\)-th particle, will be written as \(x^{(k)} = \{x_\alpha^{(k)}\}\), and for the initial state, the  $\alpha^\mathrm{th}$ coordinate of the initial position will be written as \(y = \{y_\alpha\}\). In other words, Latin superscripts index particles \(1,\ldots,N\) while Greek subscripts indicate their coordinates, \(1,\ldots,d\).

Suppose $N$ particles are released at the same position at $t=0$ and evolve in a potential \(U(\vec{x})\) according to Langevin dynamics. The evolution of the state probability distribution $p(\vec{x},t)$ is governed by the general Fokker-Planck equation,
\begin{align*}
\frac{\partial p(\vec{x},t)}{\partial t}=\sum_{\alpha,\beta} \frac{\partial^2[D_{\alpha \beta}(\vec{x},t) p(\vec{x},t)]}{\partial x_\alpha \partial x_\beta}  -\sum_{\alpha} \frac{\partial[ \mu_\alpha(\vec{x},t) p(\vec{x},t)]}{\partial x_\alpha},
\end{align*}
where $\mu_\alpha(\vec{x},t)$ is a drift term and $D_{\alpha \beta}(x,t)$ is the diffusion tensor. Since particles are independent and follow identical transition rules, the probability that $N$ particles starting at state $x$, end in states $x^{(1)},\ldots,x^{(N)}$ is
\begin{align}
\mathcal{T}(x^{(1)},\ldots,x^{(N)} \mid y ; t) = \prod_{k=1}^{N} p({x^{(k)} \mid y ; t})
\label{TP}
\end{align}
The retrodiction probability $\mathcal{R}(y \vert x^{(1)},\ldots,x^{(N)})$ is then the probability that the initial position of the cluster of particles was $y$ given the \(N\) observed final positions $\{ x^{(k)} \}$. 

\subsection{Retrodiction Entropy of a Gaussian process} 
Consider a process with (individual) probability distributions
\begin{align}
p(x^{(k)} \vert y ; t) = \prod_{\alpha=1}^d \frac{\exp [ ( x_\alpha^{(k)} - \lambda_\alpha(t) y_\alpha)^2/D_\alpha(t)]}{\sqrt{\pi D_\alpha(t)}}.
\label{form}
\end{align}
Here, the transition probability \( \mathcal{T}(x^{(1)},\ldots, x^{(N)}\vert y) \) is 
\begin{align}
\mathcal{T}&= \prod_{\alpha = 1}^d \frac{\exp [- \sum_{k=1}^N (x_\alpha^{(k)} - \lambda_\alpha(t) y_\alpha)^2/D_\alpha(t)]}{(\pi D_\alpha(t))^{N/2}}
\label{TProbability}
\end{align}
since all particles start at $x_\alpha$. Note that we allow the generalized diffusion and drift to be different in every dimension \(\alpha\). Suppose the prior probability for the initial position of the cluster of particles is Gaussian, centered at the origin, 
\begin{align}
P_0(y) = \prod_{\alpha=1}^d (2 \pi \sigma_\alpha^2)^{-1/2} \exp [-y_\alpha^2/(2 \sigma_\alpha^2)].
\label{PProbability}
\end{align}
The observation probability of a configuration is then
\begin{align}
P_t(x^{(1)},\ldots,x^{(N)} ; t) = \prod_{\alpha = 1}^d \left( 2 \pi^N \sigma_\alpha^2 D_\alpha(t)^N \right)^{-1/2} \nonumber
\\
\!\!\!\!\!\times \sqrt{\frac{D_\alpha(t)\kappa_\alpha(t)}{N \lambda_\alpha(t)^2}} \exp[-\frac{N}{D_\alpha(t)} \left( \langle x_\alpha^2 \rangle - \kappa_\alpha(t) \langle x_\alpha \rangle^2 \right)]
\label{NProbability}
\end{align}
where, $\kappa_\alpha(t) =[1 + D_\alpha(t)/(2 N \sigma_\alpha^2 \lambda_\alpha(t)^2)]^{-1}$
and $\langle x_\alpha^n \rangle =\sum_{k=1}^N [x_\alpha^{(k)} ]^n/N$. From this and \(\mathcal{T}\), \(P\), we can evaluate the retrodiction probability
\begin{align*}
\mathcal{R} & (y \vert x^{(1)},\ldots,x^{(N)} ; t) = \prod_{\alpha=1}^d \sqrt{\frac{N \lambda_\alpha(t)^2}{\pi D_\alpha(t) \kappa_\alpha(t)}}
\\
&\times \exp \left[ - \left( \frac{N \lambda_\alpha(t)^2}{D_\alpha(t) \kappa_\alpha(t)} \right) \left( y_\alpha - \frac{\kappa_\alpha(t)}{\lambda_\alpha(t)} \langle x_\alpha \rangle \right)^2 \right].
\end{align*}

As this is a Gaussian distribution, it is straightforward to evaluate its entropy, the retrodiction entropy,
\begin{align}
S_R = \frac{1}{2} \log\left[\left(\frac{\pi e}{N}\right)^d \prod_{\alpha=1}^d \frac{D_\alpha(t)}{\lambda_\alpha(t)^2 + D_\alpha(t)/(2 \sigma_\alpha^2 N)}\right].
\label{GaussianReconstructionEntropy}
\end{align}

Note that in the limit of \( \sigma_\alpha \rightarrow \infty \) in all directions, we obtain the case of a uniform (non-normalizable) prior over all space. In this case, or in the case that \(\sigma\)'s are finite and particles are ``scattered off'' by external forces, i.e. $\lambda_\alpha(t) \rightarrow \infty$ as $t\to\infty$, the retrodiction entropy is
\begin{align*}
S_R =(d/2) \log [\pi e D_{\mathrm{GM}}(t)/(N \lambda_{\mathrm{GM}}(t)^2)]
\end{align*}
where the subscript ``GM'' indicates a geometric mean over the different directions \(\alpha\). The individual entropies of the distributions \(\mathcal{T}\), \(P_0\), and \(P_t\) are listed in Appendix A, which also serves to verify (\ref{FundLemma}).

\subsection{Convex and concave potentials} 
\begin{figure}
\includegraphics[width=0.48\textwidth]{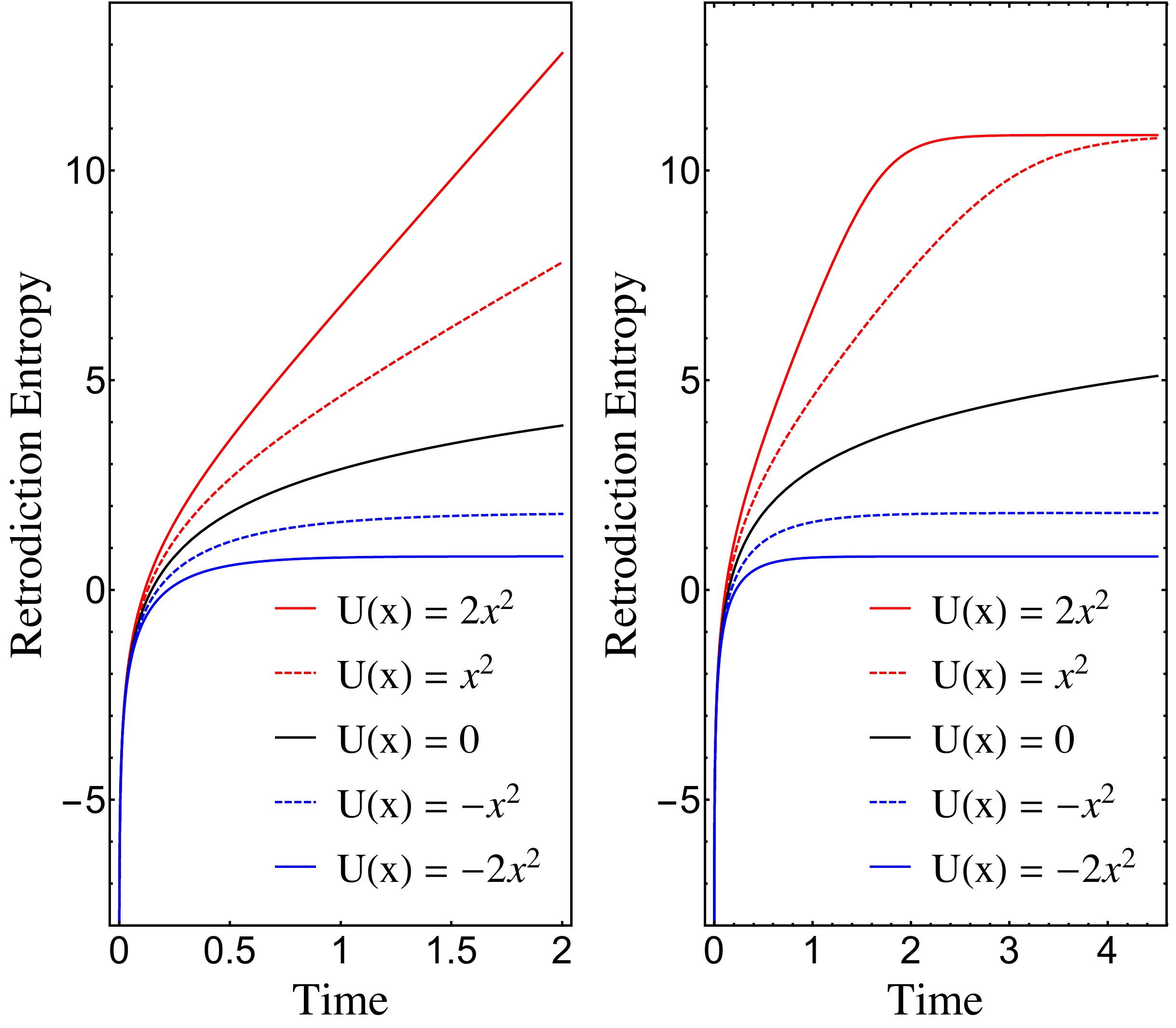}
\caption{\textbf{Retrodiction Entropy Generation.} \textbf{Left:} The retrodiction entropy $S_R$ of five particles in a convex, flat and concave potential $U(\vec{x})$ with a uniform prior. $S_R$ quantifies how poorly the initial state of the particles can be inferred, backwards in time. Free and trapped particles forget their origin monotonically, whereas particles dispersing in a concave potential remember their past no matter how much time passes. \added{\textbf{Right:} An analogous plot, but with a Gaussian prior instead of a uniform prior. Free and trapped particles saturate to having maximum retrodiction entropy, whereas particles in a concave potential still remember their past, just as in the case of a uniform prior. } }
\label{Figure1}
\end{figure}

Two processes that have analytical solutions to the Fokker-Planck equation are Wiener and Ornstein-Uhlenbeck processes, describing Brownian particles in flat $U(\vec{x})=\alpha+\vec{\beta}\cdot \vec{x}$ and parabolic $U(\vec{x}) = \alpha+\vec{\beta}\cdot \vec{x} + \theta \vec{x}^2$ potentials. We evaluate the retrodiction entropy for these special cases, and find that it diverges for particles random walking in flat and convex potentials ($\theta\geq0$) indicating that the system steadily forgets its past. In contrast, concave ($\theta\le0$) potentials have a retrodiction entropy that asymptotically approach a constant less than \(S_P\), indicating that the system always retains the memory of its initial state (see Fig.\ref{Figure1}).

\begin{figure}
\includegraphics[width=0.48\textwidth]{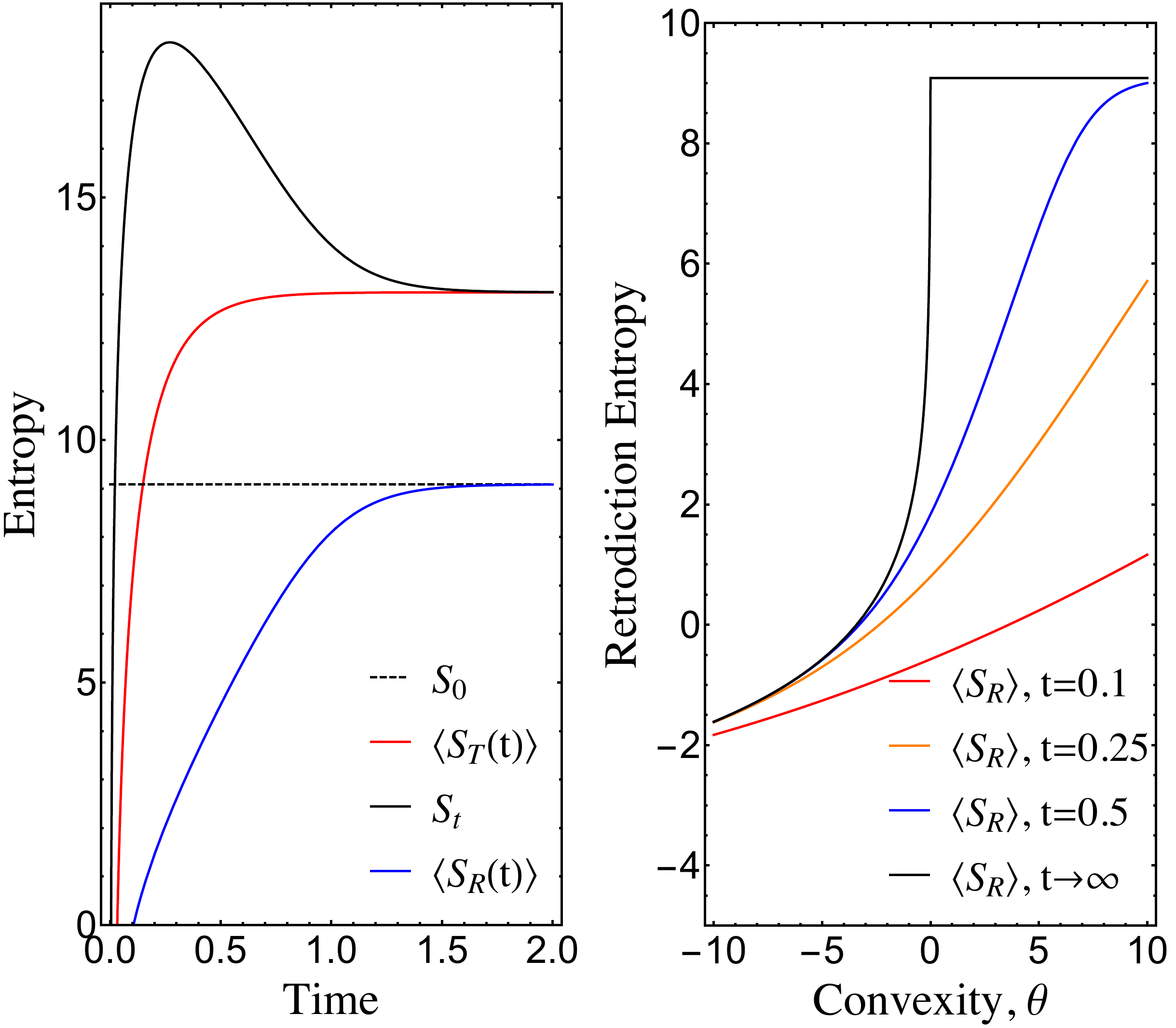}
\caption{\textbf{The Role of Convexity.} \added{\textbf{Left:} Prior entropy $S_0$, average thermodynamic entropy $\langle S_T \rangle$, and observational entropy $S_t$ for two particles in a harmonic trap, as derived in Appendix A. The retrodiction entropy is related to the other three, through \(\langle S_R \rangle = \langle S_T \rangle - (S_t - S_0)\). The prior distribution is Gaussian with \(\sigma=5\).} \textbf{Right:} \(S_R\) for the Ornstein-Uhlenbeck process at different times with a Gaussian prior, \(\sigma=5\). \added{For positive convexity, \(S_R\) converges to \(S_0\), for negative convexity, \(S_R\) converges to some smaller value, meaning some information can still be recovered.} }
\label{Figure2}
\end{figure}

The distribution of a free Brownian particle is
\begin{align*}
p(x \mid y ; t) = \prod_{\alpha=1}^d (4 \pi D_\alpha t)^{-1/2} \exp [-(x_\alpha-y_\alpha)^2/4 D_\alpha t].
\end{align*}
In this case, the functions in (\ref{form}) are \(D_\alpha(t) = 4 D_\alpha t\) and \(\lambda_\alpha(x_\alpha) = 1 \). Thus,
\begin{align*}
S_R = \frac{1}{2} \log \left[ \left(4 \pi e \right)^d \prod_{\alpha=1}^d \frac{ \sigma_\alpha^2 D_\alpha t}{2 D_\alpha t + \sigma_\alpha^2 N} \right].
\end{align*}

In the limit of \(\sigma_\alpha \rightarrow \infty\), \(S_R\) increases at a logarithmic rate at all times. If the \(\sigma\)'s are finite, then at long times,  \(S_R \rightarrow \sfrac{d}{2} \log [4 \pi e \sigma_{GM}^2]\), which is just the entropy of the prior distribution \(P_0\). For short times, we have 
\[
S_R \sim (d/2) \log \left(4 \pi e D_{GM} t/N \right)
\]

Next, we consider Brownian particles in a convex or concave harmonic potential, \(U(\vec{x}) = \theta \vec{x}^2\), described by the the Orstein-Uhlenbeck process. The probability distribution given an initial position $y$ is
\begin{align*}
p(x \mid y ; t) =\!\prod_{\alpha=1}^d \frac{\exp[ -\theta(x_\alpha-y_\alpha \mathbf{e}^{-\theta t})^2/(2 D_\alpha(1-\mathbf{e}^{-2 \theta t}))]}{\sqrt{2\theta^{-1} \pi D_\alpha (1-\mathbf{e}^{-2\theta t})}}
\end{align*}
meaning that $D_\alpha(t) = 2 D_\alpha \theta^{-1} (1-\mathbf{e}^{-2 \theta t})$ and $\lambda_\alpha(t) = e^{-\theta t}$. 
Thus,
\begin{align*}
S_R &= \frac{1}{2} \log\left[\left(2 \pi e\right)^d \prod_{\alpha=1}^d \frac{\sigma_\alpha^2 D_\alpha (1-\mathbf{e}^{-2 \theta t})}{\sigma_\alpha^2 N \theta \mathbf{e}^{-2 \theta t} + D_\alpha (1-\mathbf{e}^{-2 \theta t})}\right].
\end{align*}

In the limit of infinite \(\sigma\)'s, we get two very different long-time behaviors depending on the sign of $\theta$. For $\theta>0$ we have a harmonic trap. As $t \rightarrow \infty$, $S_R \sim d \theta t$. For $\theta<0$, we have a potential that tends to quickly force particles away from the origin. In this case, 	
\begin{align*}
S_R = (d/2)\log\left[\left(2 \pi \mathbf{e}\right) (1 - \mathbf{e}^{2 \vert \theta \vert t})D_{GM}/(N \vert \theta \vert)\right].
\end{align*}
Therefore, as $t \rightarrow \infty$, $S_R \sim \mbox{const.} - \frac{d}{2} e^{-2 \vert \theta \vert t}$. Thus, after some initial transient loss of information, our ability to reconstruct the initial state plateaus, i.e. the system always retains information about its initial state for arbitrarily long times (see Fig. \ref{Figure1}a). For finite \(\sigma\)'s, \(S_R\) has three distinct temporal regimes. It starts logarithmic, crosses over to linear, and then finally saturates to \(S_0\) (see Fig. \ref{Figure1}B, \ref{Figure2}).

\added{
In Fig. \ref{Figure1}a, we have plotted the average retrodiction entropy as a function of time for five particles in potentials with various concavities (\(\theta\) parameters, \(U(x) = \theta x^2\)). The prior is a non-normalizable uniform prior. The process is an Ornstein-Uhlenbeck process when \(\theta\ne 0\), and is the Weiner process when \(\theta = 0\). For concave potentials (in blue), the retrodiction entropy converges to a finite value. For a potential with \(\theta=0\), we recover the Wiener process, and \(S_R\) increases logarithmically. For convex potentials, \(S_R\) is asymptotically linear, diverging much more quickly than the Weiner process. }

\added{In Fig. \ref{Figure1}b, we have shown the analogous plot, but for a Gaussian prior. For concave potentials, the retrodiction entropy still saturates to a value below the prior entropy value. For convex potentials, the retrodiction entropy starts logarithmic, becomes linear, and then quickly saturates to \(S_0\). For the Wiener process, the retrodiction entropy does eventually approach the value of \(S_0\), though very slowly -- at \(t=1000\), it is still \(~2.5\%\) away from \(S_0\).
}

\added{
In Fig. \ref{Figure2}a, we show the time dependence of the entropies \(S_0\), \(\langle S_T \rangle\), \(S_t\), and \(\langle S_R \rangle\) for two particles in a convex potential with a Gaussian prior. This illustrates the fact that \(\langle S_R \rangle = \langle S_T \rangle - (S_t - S_0)\). The linear behavior of \(S_R\) in the intermediate regime can be seen before it exponentially approaches the value of the entropy of the prior, \(S_0\).}

\added{In Fig. \ref{Figure2}b, we plot the average retrodiction entropy of the Ornstein-Uhlenbeck process at specific times, starting with a Gaussian prior. In the long time limit, if the convexity is positive, the retrodiction entropy approaches the entropy of the prior distribution, \(S_0\), and hence the black line being flat for all \(\theta \ge 0\). However, if the convexity is negative (so a concave potential), we can see that the retrodiction entropy converges to a value less than \(S_0\). This indicates that by making a measurement, we gain information about the initial state of the system even after arbitrarily long times.
}

\section{\added{Retrodiction of a Chaotic System}}

\begin{figure*}
\includegraphics[width=\textwidth]{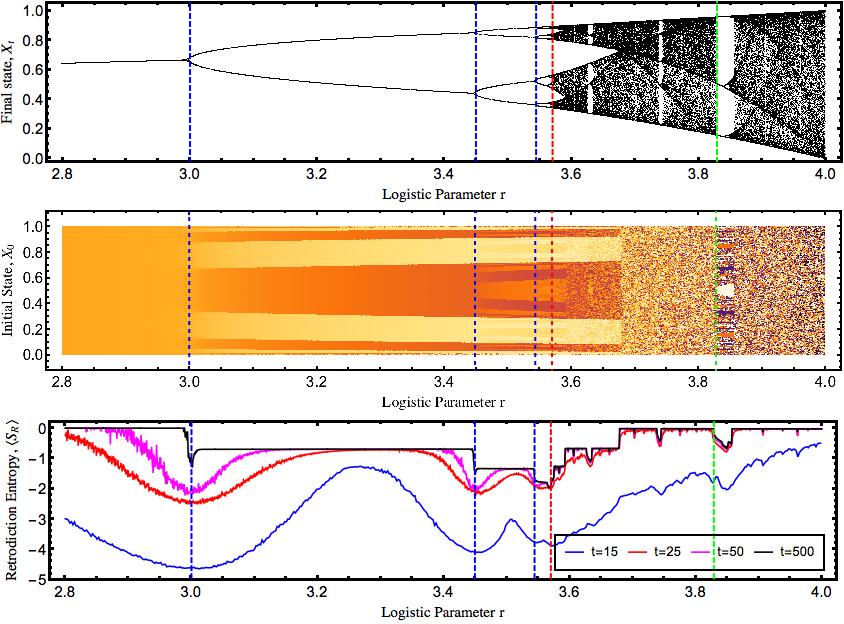}
\caption{
\added{ \textbf{Retrodiction Entropy, Bifurcations and Chaos.} The vertical dashed lines from left to right are (1) period doubling, (2) period quadrupling, (3) period $\times8$, (4) onset of chaos (red), and (5) the onset of one particular island of stability where chaos breaks off to periodic motion (green). \textbf{Top:} The bifurcation diagram for the logistic map, showing $X_t$ for multiple large $t$ values. \textbf{Middle:} Basins of attraction. The initial state $X_0$ determines the final state $X_t$, ($t=200$) within $[0,1]$ which is mapped to a color gradient from dark red (0) to light yellow (1). The change in the number of basins can be clearly seen near the vertical lines. As the system transitions into chaos, nearby points start converging to distinct final points. The system becomes chaotic (at $r\simeq 3.58$)  and then ``well mixed'' (at \(r\simeq 3.68\). \textbf{Bottom:} The (normalized) retrodiction entropy vs. logistic parameter $r$ is plotted at several different times, $t$. The black line ($t=500$) is an excellent approximation to the asymptotic limit of \(\langle S_R \rangle\). Note that in the non-chaotic regime, the retrodiction entropy converges to flat steps, whereas in the chaotic regime, the retrodiction entropy converges to steps (with values equal to that in the non-chaotic regime) with occasional dips coinciding with islands of stability. Course graining was done with \(b=500\) bins, with 10000 sample initial points per bin.} 
}
\label{ChaosMap}
\end{figure*}

\added{
To study how chaos relates to retrodictability  we consider the simplest of chaotic systems, the logistic map,
\[
X_t = r \cdot X_{t-1} (1- X_{t-1}).
\]
characterized by a single parameter $r$ which determines whether the system is chaotic. Our key result here is somewhat counter-intuitive: We find that the system is maximally retrodictable right before and right after it transitions into chaos.}

\added{The asymptotic properties of the Logistic map is well known \cite{may1976simple}. The values \(p_n\) take as $n$ tends to infinity, i.e. the attractors, is shown in the bifurcation diagram (Fig. \ref{ChaosMap}a). For small values of $r$, the trajectories are periodic. As $r$ is increased, there is a sequence of period doublings (cf. Fig. \ref{ChaosMap}, blue vertical dashes) until the system transitions to chaos at \(r\simeq 3.57\) (red vertical dashes). Within the chaotic regime, there are occasional islands of stability where periodic attractors exist. For example, at $r\simeq 3.83$ there is a period 3 attractor (green vertical dashes).
}

\added{Since the logistic map is purely deterministic, in order to define probabilities and entropies we suppose that the state of the system cannot be measured with infinite accuracy -- similar to how probability and entropy arise in classical statistical mechanics. To avoid artifacts stemming from the precise details of coarse graining, we pick very small bins with randomized positions}

\added{Specifically, we coarse grain the interval \([0,1]\) randomly into \(b\) bins by picking \(b-1\) random numbers uniformly and ordering them \(0 < x_1 < x_2 < ... < x_{b-1} < 1\). We then uniformly and randomly sample \(s\) points from each bin, and iterate each point \(\tau\) times via the logistic map. This way, we construct the probability transition matrix \(T^{(\tau)}_{j i}\), the probability that a point selected randomly from bin \(j\) ends in bin \(i\) after \(\tau\) logistic steps. Using this, and assuming a uniform prior on picking the initial point, we can obtain the retrodiction probability matrix \(R^{(\tau)}_{j i}\), and the average retrodiction entropy.}

\added{As the binning is random, the value of average retrodiction entropy is slightly different for each realization of the binning, so we average over many different random binnings. We note that we are essentially calculating the information dimension of the retrodiction probability. Information dimension \cite{grassberger1983characterization,farmer1982information} is one of several common ways to calculate fractal dimension. Our prescription here is only different in that we are applying it to our retrodiction of the original state, not to the calculation of the final state.
}

\added{Figure \ref{ChaosMap} contains several panels related to the retrodictability of the logistic map. The top panel is the bifurcation diagram for the logistic map, which we align with the other two panels to use as a reference. 
}

\added{The middle panel shows what initial states converge to what final state. Here we see the basins of attraction of the logistic map.  The vertical axis indicates the initial position of the point, whereas the color represents the value the point has after 250 iterations. We can see how the unit interval splits into domains at each bifurcation point. At the onset of chaos, even the points very near each other can end up in different phase oscillations. The degree of chaos increases several times, when sub-domains of the unit interval become more mixed. This occurs for example at \(r= 3.58\) and \(r=3.59\) before the point of complete mixing at \(r=3.68\).}

\added{The bottom panel shows the retrodiction entropy at various times. The black line, for \(t=500\) steps, is a good approximation of the asymptotic limit of \(\langle S_R \rangle\). For parameter values below the first period doubling, retrodiction entropy is at a maximum since all points in the unit interval converge to a single value, therefore observing that value does not provide any useful information about the initial state of the system. Therefore, \(S = \log V = 0\) since the ``volume'' $V$ is the unit interval. At the period doubling, the asymptotic value of \(S_R\) drops to \(-\log 2\). This is reflective of the fact that in the two period region, the measure of the set of points that converge to each period is 1/2. Therefore, the retrodiction entropy given either of the two ending positions is \(-\log 2\). This trend of reduction in average retrodiction entropy continues with every period doubling, as an equal measure of points converge to different basins.}

\added{Note that, as period doublings occur more rapidly with increasing $r$, our finite bin size prohibits us from resolving the discrete steps close to the onset of chaos. As period doublings happen exponentially quickly and exponentially close together, an exponential number of bins becomes necessary to distinguish between the entropy drops associated with successive bifurcations.}

\added{The blue vertical dashed lines in Fig. \ref{ChaosMap} show the locations of the period doublings. Near the period doubling points, there is a dramatic slowdown in convergence of \(S_R\) to its asymptotic value, which is reflective of the fact that there is a slowdown in convergence of sequences to the periodic attractor.}

\added{As period multiplicities of every power of 2 occur before the onset of chaos, the long-time limit of differential retrodiction entropy approaches negative infinity (in the limit of infinite number of bins). Even with a limited number of bins, the asymptotic retrodiction entropy hits a minimum right at the chaotic transition.}

\added{Past the point of chaos, retrodiction entropy ascends in steps with the same asymptotic values as the descending steps. The reason why the steps have the same value can be seen in the middle panel of Fig. \ref{ChaosMap}. As \(r\) approaches chaos, the system breaks the unit interval of starting positions into sub-domains that map to different periodic attractors (that are subdivided somewhat similarly to a Cantor set). After the onset of chaos, the sub-domains undergo mixing, as previously mentioned, where any point that started in that domain has an equal chance of ending up in any attractor in any sub-domain of that domain.}

\added{The reconstruction entropy in the chaotic regime also has occasional dips, which correlate with the ``islands of stability''. For example, we have marked the value \(r=3.83\) in green, which is where the logistic map has a period three oscillation. The dips around \(r=3.63\) and \(r=3.74\) occur because the logistic map is not chaotic for some values of \((x,r)\), but instead an entire neighborhood in the unit interval converges to the same attractor.}

\section{Discussion} 
The approach of using retrodiction entropy bears some similarities to other methods of inference, particularly maximum a posteriori (MAP) estimation and other Bayesian methods, but also has significant differences. Philosophically, our goal in defining $S_R$ is not to find the mode of a distribution (this is the usual goal of Bayesian inference), but to characterize the information contained in the distribution as a whole. Identifying modes, or the most likely initial state can be very misleading. For example, in highly degenerate systems, there could be many peaks in $\mathcal{R}$, each containing a small amount of probability mass. In contrast, $S_R$ characterizes the information content within the entire probability distribution.

\added{That being said, entropy does not constitute a complete characterization of a probability distribution either. For example, it might be informative to pull out a guess from \(\mathcal{R}\) and compare it with the actual initial state,
\[
\int \mathcal{R}_y(x_1) (x_1 - x_2)^2 \mathcal{R}_y(x_2) \mathbf{d}x_1 \mathbf{d}x_2.
\]
Since entropy does not take into account information about the spatial location of probability mass, it would not inform on this quantity.}

\subsection{\added{Comparison with other approaches}}
\added{
There is a long history of inference and information theory in the development of statistical mechanics. Here, we briefly review a few similar methods of doing inference and measuring predictability.}

\added{Problems in inverse statistical mechanics are generally solved by using maximum likelihood estimation (MLE) or, if prior information is available, maximum a-posterior estimation (MAP). Other methods are available, for example, the pseudolikelihood \cite{besag1974spatial}. However, most of the problems typically treated in inverse statistical physics are lattice problems, and the typical goal is to find microscopic parameters of the system given some number of (generally independent) measurements, rather than finding the state of the system in the past. 
For example, a prototypical inverse statistical mechanical problem is the inverse Ising problem \cite{nguyen2017inverse}, where the connections \(J_{ij}\) between spin variables is unknown, the spin configuration is sampled some number of times from the equilibrium distribution, and the problem is to infer the most likely matrix \(J_{ij}\).}

\added{A line of papers by J. Crutchfield and C.J. Ellison treat semi-infinite chains of random variables as consecutive states in discrete time, and suggests that the mutual information between semi-infinite sets of variables is a good measure for the amount of information about the past stored in the present \cite{crutchfield2009time,ellison2009prediction,crutchfield2010past,feldman1998discovering}. Their backwards entropy \( h_\mu = \lim_{n\to \infty} H(X_{-n+1},...,X_0)/n \) differs from our retrodiction entropy, which, in compatible notation, becomes \( \langle S_R \rangle = H(X_0) - I(X_0 ; X_t) \) (cf. (\ref{MutualInfo})). Note that while $h_\mu$ is defined for a chain of infinite time points, retrodiction entropy operates between two specific times.}

\added{The goals of computational mechanics and our retrodiction entropy approach are different. Computational mechanics asks what finite state machine can statistically reproduce a sequence or random variables. Furthermore, many of the examples they treat are not physical systems, but finite state computational processes - they look at e.g. the random insertion process \cite{crutchfield2009time}, random noisy copy, and the golden mean process \cite{ellison2009prediction}, though in \cite{feldman1998discovering} the authors look at reproducing the patterns in different Ising systems.}

\added{In addition, the constraint of having infinite pasts and futures amounts to studying systems only in equilibrium, which is not a case we would typically be interested in when studying retrodiction entropy.}

\subsection{\added{Possible generalizations}}
\added{We can loosen our formalism to make it applicable to general inference problems; not just problems in statistical mechanics. An inference problem is typically of the form where there is a space of sets of possible model parameters, \(A\), and a space of possible observed outcomes, \(\Omega\). The transition probability is the probability that an observable event occurs given a set of model parameters. There is not necessarily any variable that serves as ``time.'' As the problem is one of reconstructing parameters, and there is no time, so no ``past,'' we would call the Bayesian inverse of \(\mathcal{T}\) reconstruction probability and call the corresponding \(S_R\) reconstruction entropy (instead of retrodiction probability and entropy).}

\added{Reconstruction entropy is a measurement of how well we can determine the parameters of a system given an observed event generated from a model with unknown parameters. Retrodiction entropy is a special case of this where the set of parameters is the same as the set of observables (\(A = \Omega\)), e.g. both are phase space. Additionally, when retrodicting, we consider a parameterized family of transition probabilities, understanding this parameter to be our system time. For the more general reconstruction entropy, most of the formulas we have derived still hold, for example \cref{RetrodictionEntropy,ent,FundLemma,EquilibriumSr,DklRvsT,MutualInfo}, and the KL divergence relations in appendix B. On the other hand, results like (\ref{SNTimeInequ}) do not hold if there is no time parameter.}

\section{Conclusion}
We introduced the notion of retrodiction entropy as a measure of our ability to infer the past state of a collection of particles based on a single measurement of the system, and derived a relationship between this and thermodynamic entropy. We have established bounds on the retrodiction entropy generation rate, \added{derived a set or KL divergence relations between different relevant probabilities,} and outlined retrodiction entropy's asymptotic properties. We also showed that for systems where the initial state is an equilibrium distribution, the average forward and retrodiction entropy are identical. Lastly, we analytically solved two concrete examples, quantifying how rapidly a system of particles forgets its initial state in convex, concave and flat potentials, \added{and analyzing macrostate retrodiction entropy for a chaotic system}. Particularly, we saw that in a concave potential there is an upper limit to the loss of information pertaining the initial state, \added{and for the logistic map, we saw sharp changes in asymptotic retrodiction entropy at period doublings, and could identify islands of stability in the chaotic regime by dips in retrodiction entropy}.

The connection between thermodynamic quantities \(\langle S_T \rangle\), \(S_t\) and a purely information theoretical one, $S_R$, is in accordance with the seminal works of Maxwell, Smoluchowski, Landauer, Szillard, Beckenstein, and others \cite{jaynes1957informationA,jaynes1957informationB,leff2014maxwell,shannon1998mathematical}. We now know, from (\ref{FundLemma}), that thermodynamic entropy at present time not only quantifies the information content of the state of the system at present time, it also relates to how precisely information about \added{the original state of the system can be recovered after some amount of time has passed.}


\section*{Appendix A: Derivation of the Relationship between Retrodiction Entropy and Thermodynamic Entropy}
We use sum notation throughout, although these could be replaced with integrals. Suppose $P$ is normalized. Then (\ref{FundLemma}) can be proved through simple integration:
\begin{align*}
\langle S_R \rangle = &\sum_\omega P_t(\omega) S_R(\xi)= -\sum_\omega P_t(\omega) \sum_\alpha \mathcal{R}_\omega (\alpha) \log \mathcal{R}_\omega(\alpha)
\\
= -&\sum_{\omega,\alpha}\mathcal{T}_\alpha(\omega) P_0(\alpha) \log \left( \frac{\mathcal{T}_\alpha(\omega) P_0(\alpha)}{P_t(\omega)} \right)
\\
= -&\sum_{\omega,\alpha}P_0(\alpha)\mathcal{T}_\alpha(\omega) \log \mathcal{T}_\alpha(\omega) + \sum_\omega P_t(\omega) \log P_t(\omega) \\
-&\sum_\alpha P_0(\alpha) \log P_0(\alpha) = \langle S_T \rangle - (S_t - S_0).
\end{align*}
where we substituted \(P_t(\omega) = \sum_\alpha \mathcal{T}_\alpha(\omega) P_0(\alpha) \).

As an explicit example of this, consider the Gaussian process family we discussed in the paper, with \(\mathcal{T}, P_0, P_t\) given by (\ref{TProbability}), (\ref{PProbability}) and (\ref{NProbability}). For this case,
\begin{align*}
S_T &= \langle S_T \rangle = \frac{1}{2} \log \left[ \pi^N \prod_{\alpha=1}^d D_\alpha(t)^N \right] + \frac{N d}{2}
\\
S_0 &= \frac{1}{2} \log \left[ (2 \pi e)^d \prod_{\alpha=1}^d \sigma_\alpha^2 \right]
\\
S_t &= \frac{1}{2} \log \left[ (2 \pi^N N)^d \prod_{\alpha=1}^d \sigma_\alpha^2 D_\alpha(t)^N \frac{\lambda_\alpha(t)^2}{D_\alpha(t) \kappa_\alpha(t)} \right] + \frac{N d}{2}
\end{align*}
from which it can be shown, using (\ref{FundLemma}), that
\begin{align*}
S_R = \langle S_R \rangle = \frac{1}{2} \log \left[\left( \frac{\pi e}{N} \right)^d \prod_{\alpha=1}^d \frac{D_\alpha(t) \kappa_\alpha(t)}{\lambda_\alpha(t)^2}\right]
\end{align*}

\section*{Appendix B: KL-Divergence Relations}
Here, we derive (\ref{DklRvsT}). We start with the definition of KL-divergence:
\begin{align*}
&\KL{\mathcal{R}_{\omega_1}}{\mathcal{R}_{\omega_2}} = - \sum_\alpha \mathcal{R}_{\omega_1} (\alpha)\log \frac{\mathcal{R}_{\omega_2} (\alpha)}{\mathcal{R}_{\omega_1} (\alpha)}
\\
&= -\sum_\xi \frac{\mathcal{T}_\alpha(\omega_1) P_0(\alpha)}{P_t(\omega_1)} \bigg[ \log \left( \frac{\mathcal{T}_\alpha(\omega_2) }{\mathcal{T}_\alpha(\omega_1) } \right)+ \log \left( \frac{P_t(\omega_1)}{P_t(\omega_2)} \right) \bigg].
\end{align*}
Averaging over \(\omega\)'s with the probability weight \(P_t(\omega_1) P_t(\omega_2)\), the first term in the brackets gives
\begin{align*}
& -\sum_{\omega_1,\omega_2,\alpha} P_t(\omega_2) \mathcal{T}_\alpha(\omega_1) P_0(\alpha) \log[\mathcal{T}_\alpha(\omega_2) /\mathcal{T}_\alpha(\omega_1)]
\\
&= -\sum_{\omega_1,\omega_2,\alpha} P_0(\alpha) P_t(\omega_2) \left[ \mathcal{T}_\alpha(\omega_1) \log \mathcal{T}_\alpha(\omega_2) - \mathcal{T}_\alpha(\omega_1) \log \mathcal{T}_\alpha(\omega_1) \right]
\\
&= -\sum_{\omega_1,\omega_2,\alpha} P_0(\alpha) P_t(\omega_2)  \log \mathcal{T}_\alpha(\omega_2) - P_0(\alpha) \mathcal{T}_\alpha(\omega_1) \log \mathcal{T}_\alpha(\omega_1)
\\
&= - \langle S_T \rangle - \sum_{\omega_2} P_t(\omega_1) \log \mathcal{T}_\alpha(\omega_2)
= S_t - \langle S_T \rangle + \sum_\alpha \KL{P_t}{\mathcal{T}_\alpha}
\end{align*}
(we have used the fact that \(\KL{A}{B} =- \sum A \log B - S_A \) and \(\sum_\omega \mathcal{T}_\alpha(\omega) = 1\)) whereas the second term gives
\begin{align*}
& -\sum_{\omega_1,\omega_2,\alpha} P_t(\omega_2) \mathcal{T}_\alpha(\omega_1) P_0(\alpha) \log \frac{P_t(\omega_1)}{P_t(\omega_2)}
\\
& =-\sum_{\omega_1,\omega_2} P_t(\omega_2) \left( \sum_\alpha \mathcal{T}_\alpha(\omega_1) P_0(\alpha) \right) \log \frac{P_t(\omega_1)}{P_t(\omega_2)}
\\
&= -\sum_{\omega_1, \omega_2} P_t(\omega_2) P_t(\omega_1) \log \frac{P_t(\omega_1)}{P_t(\omega_2)} 
\\
&= S_t - S_t= 0.
\end{align*}
Putting everything together,
\begin{align*}
\langle \KL{\mathcal{R}_{\omega_1}}{\mathcal{R}_{\omega_2}} \rangle = \langle \KL{\mathcal{N}}{\mathcal{T}_\xi} \rangle + S_t - \langle S_T \rangle.
\end{align*}
The second term here is,
\begin{align*}
\langle \KL{\mathcal{N}}{\mathcal{T}_\xi} \rangle &= -\sum_{\xi, \omega} P(\xi) \mathcal{N}(\omega) \log \frac{\mathcal{T}_\xi(\omega)}{\mathcal{N}(\omega)}
\\
&= -\sum_{\xi_1, \xi_2} P(\xi_1) P(\xi_2) \mathcal{T}_{\xi_1}(\omega) \log \mathcal{T}_{\xi_2}(\omega) - S_t
\\
&= \langle \KL{\mathcal{T}_{\xi_1}}{\mathcal{T}_{\xi_2}} \rangle + \langle S_T \rangle - S_t.
\end{align*}
Putting these equations together gives us Eq. (\ref{DklRvsT}).

\added{We can take the KL divergence between any pair of distributions that have a common domain. It is natural to only compare distributions that are either both on the final state or both on the initial state. Furthermore, as the KL divergence is asymmetric, we can ask about both orderings. The six options are \((\mathcal{T},\mathcal{T})\), \((\mathcal{T},P_t)\), \((P_t,\mathcal{T})\), \((\mathcal{R},\mathcal{R})\), \((\mathcal{R},P_0)\), and \((P_0,\mathcal{R})\). In a similar way to our derivations above, we can find relations between the averages of the KL divergence between all these pair in terms of each other or in terms of entropies:}
\begin{align*}
\langle \KL{\mathcal{T}_{\alpha_1}}{\mathcal{T}_{\alpha_2}} \rangle &= \langle \KL{P_0}{\mathcal{R}_{\omega}} \rangle + S_t - \langle S_T \rangle
\\
\langle \KL{\mathcal{T}_{\alpha}}{P_t} \rangle &= \langle \KL{\mathcal{R}_{\omega}}{P_0} \rangle = S_0 - \langle S_R \rangle = S_t - \langle S_T \rangle 
\\
\langle \KL{P_t}{\mathcal{T}_{\alpha}} \rangle &= \langle \KL{P_0}{\mathcal{R}_{\omega}} \rangle
\\
\langle \KL{\mathcal{T}_{\alpha_1}}{\mathcal{T}_{\alpha_2}} \rangle &= \langle \KL{\mathcal{R}_{\omega_1}}{\mathcal{R}_{\omega_2}} \rangle
\end{align*}
\added{One can put these together to derive relations for the averages of the symmetric combinations of KL divergences.}
\begin{align*}
\left\langle \KL{\mathcal{T}_{\alpha}}{P_t} + \KL{P_t}{\mathcal{T}_{\alpha}} \right\rangle &= \langle \KL{\mathcal{T}_{\alpha_1}}{\mathcal{T}_{\alpha_2}} \rangle
\\
\left\langle \KL{\mathcal{R}_{\omega}}{P_0} + \KL{P_0}{\mathcal{R}_{\omega}} \right\rangle &= \langle \KL{\mathcal{T}_{\alpha_1}}{\mathcal{T}_{\alpha_2}} \rangle
\end{align*}
\section*{Appendix C: Limits on the size of observational and retrodiction entropy}

We can use Jensen's inequality to put an upper bound on the time rate of change of \(S_t\). Since \(-\log x\) is a convex function, we have the inequality,
\begin{align*}
-\log\sum_\omega  P_0(\alpha) \mathcal{T}_\alpha(\omega)\le -\sum_\alpha P_0(\alpha) \log \mathcal{T}_\alpha(\omega).
\end{align*}
Start with the definition of \(S_t\), then apply Jensen's inequality:
\begin{align*}
\dot{S}_t &=\!-\!\sum_\xi \dot{P}_t(\omega) \log P_t(\omega)=\!-\!\sum_\omega \dot{P}_t(\omega) \log\sum_\alpha P_0(\alpha) \mathcal{T}_\alpha(\omega) 
\\
&\le -\sum_{\alpha,\omega} P_0(\alpha) \dot{P}_t(\omega) \log \mathcal{T}_\alpha(\omega)
\\
&= - \sum_{\alpha,\omega} P_0(\alpha) \dot{P}_t(\omega) \left( \log \frac{\mathcal{T}_\alpha(\omega)}{P_t(\omega)} + \log P_t(\omega) \right)
\\
&= -\sum_{\alpha,\omega} P_0(\alpha)\dot{P}_t(\omega) \log \frac{\mathcal{T}_\alpha(\omega)}{P_t(\omega)} + \dot{S}_t.
\end{align*}
Canceling the \(\dot{S}_N\) terms on both sides yields 
\[
0 \le -\sum_\alpha P_0(\alpha) \sum_\omega \dot{P}_t(\omega) \log \frac{\mathcal{T}_\alpha(\omega)}{P_t(\omega)}
\]
which bears some similarity to the KL-divergence. The derivative of an arbitrary KL-divergence is
\begin{align*}
\frac{\partial}{\partial t} \KL{p}{q} = -\sum \dot{p} \log \frac{q}{p} - \sum \frac{p}{q} \dot{q}.
\end{align*}
Using this in the preceding inequality, we get
\begin{align*}
0 &\le \frac{\partial}{\partial t} \langle \KL{P_t}{\mathcal{T}_\alpha} \rangle + \sum_\alpha P_0(\alpha) \sum_\omega P_t(\omega) \frac{\partial}{\partial t} \log \mathcal{T}_\alpha(\omega)
\\
&= \frac{\partial}{\partial t} \langle \KL{P_t}{\mathcal{T}_\alpha} \rangle + \sum_\alpha P_0(\alpha) \sum_\omega P_t(\omega) \frac{\partial}{\partial t} \log \frac{\mathcal{R}_\omega(\alpha) P_t(\omega)}{P_0(\alpha)}
\\
&= \frac{\partial}{\partial t} \langle \KL{P_t}{\mathcal{T}_\alpha} \rangle + \sum_\omega P_t(\omega) \sum_\alpha P_0(\alpha) \frac{\partial}{\partial t} \log \frac{\mathcal{R}_\omega(\alpha)}{P_0(\alpha)}
\\
&= \frac{\partial}{\partial t} \langle \KL{P_t}{\mathcal{T}_\alpha} \rangle - \langle \frac{\partial}{\partial t} \KL{P_0}{\mathcal{R}_\omega} \rangle.
\end{align*}
Using the expression we previously discussed for \(\langle \KL{P_t}{\mathcal{T}_\alpha} \rangle\), we can reintroduce \(\dot{S}_t\) to the equation,
\[
\dot{S}_t \le \langle \dot{S}_T \rangle + \frac{\partial}{\partial t} \langle \KL{\mathcal{T}_{\alpha_1}}{\mathcal{T}_{\alpha_2}} \rangle - \langle \frac{\partial}{\partial t} \KL{P_0}{\mathcal{R}_\omega} \rangle.
\]
We can also write this as a lower bound on \(\frac{\partial}{\partial t} \langle S_R \rangle \) via (\ref{FundLemma})
\begin{align}
\frac{\partial}{\partial t} \langle S_R \rangle \ge -\frac{\partial}{\partial t} \langle \KL{\mathcal{T}_{\alpha_1}}{\mathcal{T}_{\alpha_2}} \rangle + \langle \frac{\partial}{\partial t} \KL{P_0}{\mathcal{R}_\omega} \rangle
\label{SRDotInequality}
\end{align}
Now we will make use of the fact that for a Markov process, the relative entropy of two distributions is non-increasing \cite{cover2012elements}. We include this theorem below for the sake of completeness.

\textbf{Theorem:} Consider two probability distributions \(p\), \(q\), on the same state space. Then at any times \(t_1 < t_2\), 
\[
\KL{p_{t_1}}{q_{t_1}} \ge \KL{p_{t_2}}{q_{t_2}}
\]

\emph{Proof:} Let \(s<t\). Then,
\begin{align*}
& \KL{p(x_t \vert x_s)}{q(x_t \vert x_s)}
\\
&= \KL{p(x_t)}{q(x_t)} + \KL{p(x_s \vert x_t)}{q(x_s \vert x_t)}
\\
&= \KL{p(x_s)}{q(x_s)} + \KL{p(x_t \vert x_s)}{q(x_t \vert x_s)}
\end{align*}

By the definition of Markov, \( p(x_t \vert x_s) = q(x_t \vert x_s) \), so \(\KL{p(x_t \vert x_s)}{q(x_t \vert x_s)}\) = 0. Then, subtracting the second and third lines, we get
\[
\KL{p_t}{q_t} - \KL{p_s}{q_s} = -\KL{p_{s,t}}{q_{s,t}} \le 0 \qquad\qed
\]

If our forward dynamics are Markovian (as they are, for example, in the case of diffusion), this theorem holds and \(\frac{\partial}{\partial t}\KL{\mathcal{T}_{\alpha_1}}{\mathcal{T}_{\alpha_2}} \le 0\) for all \(\alpha_1\), \(\alpha_2\). Therefore, the first term on the right hand side of eq. (\ref{SRDotInequality}) is non-negative.

The second term of eq. (\ref{SRDotInequality}) is harder to work with. Intuitively, we expect \(\mathcal{R}\) to approach \(P\) as we lose information about the past due to stochastic events. So we expect \(\KL{P}{\mathcal{R}_\xi}\) to eventually reach a minimum for any fixed \(\xi\). As long as \(\langle \KL{P}{\mathcal{R}_\xi} \rangle\) decreases more slowly than \( \langle \KL{\mathcal{T}_{\omega_1}}{\mathcal{T}_{\omega_2}}\), this bound is good enough to guarantee that
\(\partial\langle S_R \rangle/\partial t\ge 0\).

\bibliographystyle{apsrev4-1}
\bibliography{bibliography.bib}

\begin{thebibliography}{30}%
\makeatletter
\providecommand \@ifxundefined [1]{%
 \@ifx{#1\undefined}
}%
\providecommand \@ifnum [1]{%
 \ifnum #1\expandafter \@firstoftwo
 \else \expandafter \@secondoftwo
 \fi
}%
\providecommand \@ifx [1]{%
 \ifx #1\expandafter \@firstoftwo
 \else \expandafter \@secondoftwo
 \fi
}%
\providecommand \natexlab [1]{#1}%
\providecommand \enquote  [1]{``#1''}%
\providecommand \bibnamefont  [1]{#1}%
\providecommand \bibfnamefont [1]{#1}%
\providecommand \citenamefont [1]{#1}%
\providecommand \href@noop [0]{\@secondoftwo}%
\providecommand \href [0]{\begingroup \@sanitize@url \@href}%
\providecommand \@href[1]{\@@startlink{#1}\@@href}%
\providecommand \@@href[1]{\endgroup#1\@@endlink}%
\providecommand \@sanitize@url [0]{\catcode `\\12\catcode `\$12\catcode
  `\&12\catcode `\#12\catcode `\^12\catcode `\_12\catcode `\%12\relax}%
\providecommand \@@startlink[1]{}%
\providecommand \@@endlink[0]{}%
\providecommand \url  [0]{\begingroup\@sanitize@url \@url }%
\providecommand \@url [1]{\endgroup\@href {#1}{\urlprefix }}%
\providecommand \urlprefix  [0]{URL }%
\providecommand \Eprint [0]{\href }%
\providecommand \doibase [0]{http://dx.doi.org/}%
\providecommand \selectlanguage [0]{\@gobble}%
\providecommand \bibinfo  [0]{\@secondoftwo}%
\providecommand \bibfield  [0]{\@secondoftwo}%
\providecommand \translation [1]{[#1]}%
\providecommand \BibitemOpen [0]{}%
\providecommand \bibitemStop [0]{}%
\providecommand \bibitemNoStop [0]{.\EOS\space}%
\providecommand \EOS [0]{\spacefactor3000\relax}%
\providecommand \BibitemShut  [1]{\csname bibitem#1\endcsname}%
\let\auto@bib@innerbib\@empty
\bibitem [{\citenamefont
  {Jaynes}(1957{\natexlab{a}})}]{jaynes1957informationA}%
  \BibitemOpen
  \bibfield  {author} {\bibinfo {author} {\bibfnamefont {E.~T.}\ \bibnamefont
  {Jaynes}},\ }\href@noop {} {\bibfield  {journal} {\bibinfo  {journal}
  {Physical review}\ }\textbf {\bibinfo {volume} {106}},\ \bibinfo {pages}
  {620} (\bibinfo {year} {1957}{\natexlab{a}})}\BibitemShut {NoStop}%
\bibitem [{\citenamefont
  {Jaynes}(1957{\natexlab{b}})}]{jaynes1957informationB}%
  \BibitemOpen
  \bibfield  {author} {\bibinfo {author} {\bibfnamefont {E.~T.}\ \bibnamefont
  {Jaynes}},\ }\href@noop {} {\bibfield  {journal} {\bibinfo  {journal}
  {Physical review}\ }\textbf {\bibinfo {volume} {108}},\ \bibinfo {pages}
  {171} (\bibinfo {year} {1957}{\natexlab{b}})}\BibitemShut {NoStop}%
\bibitem [{\citenamefont {Shannon}\ and\ \citenamefont
  {Weaver}(1998)}]{shannon1998mathematical}%
  \BibitemOpen
  \bibfield  {author} {\bibinfo {author} {\bibfnamefont {C.~E.}\ \bibnamefont
  {Shannon}}\ and\ \bibinfo {author} {\bibfnamefont {W.}~\bibnamefont
  {Weaver}},\ }\href@noop {} {\emph {\bibinfo {title} {The mathematical theory
  of communication}}}\ (\bibinfo  {publisher} {University of Illinois press},\
  \bibinfo {year} {1998})\BibitemShut {NoStop}%
\bibitem [{\citenamefont {Leff}\ and\ \citenamefont
  {Rex}(2014)}]{leff2014maxwell}%
  \BibitemOpen
  \bibfield  {author} {\bibinfo {author} {\bibfnamefont {H.~S.}\ \bibnamefont
  {Leff}}\ and\ \bibinfo {author} {\bibfnamefont {A.~F.}\ \bibnamefont {Rex}},\
  }\href@noop {} {\emph {\bibinfo {title} {Maxwell's demon: entropy,
  information, computing}}}\ (\bibinfo  {publisher} {Princeton University
  Press},\ \bibinfo {year} {2014})\BibitemShut {NoStop}%
\bibitem [{\citenamefont {Box}\ and\ \citenamefont
  {Tiao}(2011)}]{box2011bayesian}%
  \BibitemOpen
  \bibfield  {author} {\bibinfo {author} {\bibfnamefont {G.~E.}\ \bibnamefont
  {Box}}\ and\ \bibinfo {author} {\bibfnamefont {G.~C.}\ \bibnamefont {Tiao}},\
  }\href@noop {} {\emph {\bibinfo {title} {Bayesian inference in statistical
  analysis}}},\ Vol.~\bibinfo {volume} {40}\ (\bibinfo  {publisher} {John Wiley
  \& Sons},\ \bibinfo {year} {2011})\BibitemShut {NoStop}%
\bibitem [{\citenamefont {Welling}\ and\ \citenamefont
  {Teh}(2011)}]{welling2011bayesian}%
  \BibitemOpen
  \bibfield  {author} {\bibinfo {author} {\bibfnamefont {M.}~\bibnamefont
  {Welling}}\ and\ \bibinfo {author} {\bibfnamefont {Y.~W.}\ \bibnamefont
  {Teh}},\ }in\ \href@noop {} {\emph {\bibinfo {booktitle} {Proceedings of the
  28th International Conference on Machine Learning (ICML-11)}}}\ (\bibinfo
  {year} {2011})\ pp.\ \bibinfo {pages} {681--688}\BibitemShut {NoStop}%
\bibitem [{\citenamefont {Desmarais}\ and\ \citenamefont
  {Cranmer}(2012)}]{desmarais2012statistical}%
  \BibitemOpen
  \bibfield  {author} {\bibinfo {author} {\bibfnamefont {B.~A.}\ \bibnamefont
  {Desmarais}}\ and\ \bibinfo {author} {\bibfnamefont {S.~J.}\ \bibnamefont
  {Cranmer}},\ }\href@noop {} {\bibfield  {journal} {\bibinfo  {journal}
  {Physica A: Statistical Mechanics and its Applications}\ }\textbf {\bibinfo
  {volume} {391}},\ \bibinfo {pages} {1865} (\bibinfo {year}
  {2012})}\BibitemShut {NoStop}%
\bibitem [{\citenamefont {Nguyen}\ and\ \citenamefont
  {Vural}(2017)}]{nguyen2017morphological}%
  \BibitemOpen
  \bibfield  {author} {\bibinfo {author} {\bibfnamefont {V.~A.~T.}\
  \bibnamefont {Nguyen}}\ and\ \bibinfo {author} {\bibfnamefont {D.~C.}\
  \bibnamefont {Vural}},\ }\href@noop {} {\bibfield  {journal} {\bibinfo
  {journal} {Physical Review E}\ }\textbf {\bibinfo {volume} {96}},\ \bibinfo
  {pages} {032314} (\bibinfo {year} {2017})}\BibitemShut {NoStop}%
\bibitem [{\citenamefont {Hansen}\ \emph {et~al.}(2006)\citenamefont {Hansen},
  \citenamefont {Nagy},\ and\ \citenamefont {O'leary}}]{hansen2006deblurring}%
  \BibitemOpen
  \bibfield  {author} {\bibinfo {author} {\bibfnamefont {P.~C.}\ \bibnamefont
  {Hansen}}, \bibinfo {author} {\bibfnamefont {J.~G.}\ \bibnamefont {Nagy}}, \
  and\ \bibinfo {author} {\bibfnamefont {D.~P.}\ \bibnamefont {O'leary}},\
  }\href@noop {} {\emph {\bibinfo {title} {Deblurring images: matrices,
  spectra, and filtering}}}\ (\bibinfo  {publisher} {SIAM},\ \bibinfo {year}
  {2006})\BibitemShut {NoStop}%
\bibitem [{\citenamefont {Chan}\ and\ \citenamefont
  {Chen}(2010)}]{chan2010multilevel}%
  \BibitemOpen
  \bibfield  {author} {\bibinfo {author} {\bibfnamefont {R.~H.}\ \bibnamefont
  {Chan}}\ and\ \bibinfo {author} {\bibfnamefont {K.}~\bibnamefont {Chen}},\
  }\href@noop {} {\bibfield  {journal} {\bibinfo  {journal} {SIAM Journal on
  Scientific Computing}\ }\textbf {\bibinfo {volume} {32}},\ \bibinfo {pages}
  {1043} (\bibinfo {year} {2010})}\BibitemShut {NoStop}%
\bibitem [{\citenamefont {Ullersma}(1966)}]{ullersma1966exactly}%
  \BibitemOpen
  \bibfield  {author} {\bibinfo {author} {\bibfnamefont {P.}~\bibnamefont
  {Ullersma}},\ }\href@noop {} {\bibfield  {journal} {\bibinfo  {journal}
  {Physica}\ }\textbf {\bibinfo {volume} {32}},\ \bibinfo {pages} {27}
  (\bibinfo {year} {1966})}\BibitemShut {NoStop}%
\bibitem [{\citenamefont {Yu}\ \emph {et~al.}(2015)\citenamefont {Yu},
  \citenamefont {Eckmann}, \citenamefont {Ayyaswamy},\ and\ \citenamefont
  {Radhakrishnan}}]{yu2015composite}%
  \BibitemOpen
  \bibfield  {author} {\bibinfo {author} {\bibfnamefont {H.-Y.}\ \bibnamefont
  {Yu}}, \bibinfo {author} {\bibfnamefont {D.~M.}\ \bibnamefont {Eckmann}},
  \bibinfo {author} {\bibfnamefont {P.~S.}\ \bibnamefont {Ayyaswamy}}, \ and\
  \bibinfo {author} {\bibfnamefont {R.}~\bibnamefont {Radhakrishnan}},\
  }\href@noop {} {\bibfield  {journal} {\bibinfo  {journal} {Physical Review
  E}\ }\textbf {\bibinfo {volume} {91}},\ \bibinfo {pages} {052303} (\bibinfo
  {year} {2015})}\BibitemShut {NoStop}%
\bibitem [{\citenamefont {Coffey}\ and\ \citenamefont
  {Kalmykov}(2012)}]{coffey2012langevin}%
  \BibitemOpen
  \bibfield  {author} {\bibinfo {author} {\bibfnamefont {W.~T.}\ \bibnamefont
  {Coffey}}\ and\ \bibinfo {author} {\bibfnamefont {Y.~P.}\ \bibnamefont
  {Kalmykov}},\ }\href@noop {} {\emph {\bibinfo {title} {The Langevin equation:
  with applications to stochastic problems in physics, chemistry and electrical
  engineering}}},\ Vol.~\bibinfo {volume} {27}\ (\bibinfo  {publisher} {World
  Scientific},\ \bibinfo {year} {2012})\BibitemShut {NoStop}%
\bibitem [{\citenamefont {Wolf}(1988)}]{wolf1988lie}%
  \BibitemOpen
  \bibfield  {author} {\bibinfo {author} {\bibfnamefont {F.}~\bibnamefont
  {Wolf}},\ }\href@noop {} {\bibfield  {journal} {\bibinfo  {journal} {Journal
  of mathematical physics}\ }\textbf {\bibinfo {volume} {29}},\ \bibinfo
  {pages} {305} (\bibinfo {year} {1988})}\BibitemShut {NoStop}%
\bibitem [{\citenamefont {Hashemi}(2015)}]{hashemi2015group}%
  \BibitemOpen
  \bibfield  {author} {\bibinfo {author} {\bibfnamefont {M.}~\bibnamefont
  {Hashemi}},\ }\href@noop {} {\bibfield  {journal} {\bibinfo  {journal}
  {Physica A: Statistical Mechanics and its Applications}\ }\textbf {\bibinfo
  {volume} {417}},\ \bibinfo {pages} {141} (\bibinfo {year}
  {2015})}\BibitemShut {NoStop}%
\bibitem [{\citenamefont {Bernstein}\ and\ \citenamefont
  {Brown}(1984)}]{bernstein1984supersymmetry}%
  \BibitemOpen
  \bibfield  {author} {\bibinfo {author} {\bibfnamefont {M.}~\bibnamefont
  {Bernstein}}\ and\ \bibinfo {author} {\bibfnamefont {L.~S.}\ \bibnamefont
  {Brown}},\ }\href@noop {} {\bibfield  {journal} {\bibinfo  {journal}
  {Physical review letters}\ }\textbf {\bibinfo {volume} {52}},\ \bibinfo
  {pages} {1933} (\bibinfo {year} {1984})}\BibitemShut {NoStop}%
\bibitem [{\citenamefont {Carrillo}\ and\ \citenamefont
  {Toscani}(1998)}]{carrillo1998exponential}%
  \BibitemOpen
  \bibfield  {author} {\bibinfo {author} {\bibfnamefont {J.~A.}\ \bibnamefont
  {Carrillo}}\ and\ \bibinfo {author} {\bibfnamefont {G.}~\bibnamefont
  {Toscani}},\ }\href@noop {} {\bibfield  {journal} {\bibinfo  {journal}
  {Mathematical methods in the applied sciences}\ }\textbf {\bibinfo {volume}
  {21}},\ \bibinfo {pages} {1269} (\bibinfo {year} {1998})}\BibitemShut
  {NoStop}%
\bibitem [{\citenamefont {Toscani}(1999)}]{toscani1999entropy}%
  \BibitemOpen
  \bibfield  {author} {\bibinfo {author} {\bibfnamefont {G.}~\bibnamefont
  {Toscani}},\ }\href@noop {} {\bibfield  {journal} {\bibinfo  {journal}
  {Quarterly of Applied Mathematics}\ }\textbf {\bibinfo {volume} {57}},\
  \bibinfo {pages} {521} (\bibinfo {year} {1999})}\BibitemShut {NoStop}%
\bibitem [{\citenamefont {Schw{\"a}mmle}\ \emph {et~al.}(2007)\citenamefont
  {Schw{\"a}mmle}, \citenamefont {Curado},\ and\ \citenamefont
  {Nobre}}]{schwammle2007general}%
  \BibitemOpen
  \bibfield  {author} {\bibinfo {author} {\bibfnamefont {V.}~\bibnamefont
  {Schw{\"a}mmle}}, \bibinfo {author} {\bibfnamefont {E.~M.}\ \bibnamefont
  {Curado}}, \ and\ \bibinfo {author} {\bibfnamefont {F.~D.}\ \bibnamefont
  {Nobre}},\ }\href@noop {} {\bibfield  {journal} {\bibinfo  {journal} {The
  European Physical Journal B-Condensed Matter and Complex Systems}\ }\textbf
  {\bibinfo {volume} {58}},\ \bibinfo {pages} {159} (\bibinfo {year}
  {2007})}\BibitemShut {NoStop}%
\bibitem [{\citenamefont {Plastino}\ \emph {et~al.}(1997)\citenamefont
  {Plastino}, \citenamefont {Miller},\ and\ \citenamefont
  {Plastino}}]{plastino1997minimum}%
  \BibitemOpen
  \bibfield  {author} {\bibinfo {author} {\bibfnamefont {A.~R.}\ \bibnamefont
  {Plastino}}, \bibinfo {author} {\bibfnamefont {H.~G.}\ \bibnamefont
  {Miller}}, \ and\ \bibinfo {author} {\bibfnamefont {A.}~\bibnamefont
  {Plastino}},\ }\href@noop {} {\bibfield  {journal} {\bibinfo  {journal}
  {Physical Review E}\ }\textbf {\bibinfo {volume} {56}},\ \bibinfo {pages}
  {3927} (\bibinfo {year} {1997})}\BibitemShut {NoStop}%
\bibitem [{\citenamefont {Cover}\ and\ \citenamefont
  {Thomas}(2012)}]{cover2012elements}%
  \BibitemOpen
  \bibfield  {author} {\bibinfo {author} {\bibfnamefont {T.~M.}\ \bibnamefont
  {Cover}}\ and\ \bibinfo {author} {\bibfnamefont {J.~A.}\ \bibnamefont
  {Thomas}},\ }\href@noop {} {\emph {\bibinfo {title} {Elements of information
  theory}}}\ (\bibinfo  {publisher} {John Wiley \& Sons},\ \bibinfo {year}
  {2012})\BibitemShut {NoStop}%
\bibitem [{\citenamefont {May}(1976)}]{may1976simple}%
  \BibitemOpen
  \bibfield  {author} {\bibinfo {author} {\bibfnamefont {R.~M.}\ \bibnamefont
  {May}},\ }\href@noop {} {\bibfield  {journal} {\bibinfo  {journal} {Nature}\
  }\textbf {\bibinfo {volume} {261}},\ \bibinfo {pages} {459} (\bibinfo {year}
  {1976})}\BibitemShut {NoStop}%
\bibitem [{\citenamefont {Grassberger}\ and\ \citenamefont
  {Procaccia}(1983)}]{grassberger1983characterization}%
  \BibitemOpen
  \bibfield  {author} {\bibinfo {author} {\bibfnamefont {P.}~\bibnamefont
  {Grassberger}}\ and\ \bibinfo {author} {\bibfnamefont {I.}~\bibnamefont
  {Procaccia}},\ }\href@noop {} {\bibfield  {journal} {\bibinfo  {journal}
  {Physical review letters}\ }\textbf {\bibinfo {volume} {50}},\ \bibinfo
  {pages} {346} (\bibinfo {year} {1983})}\BibitemShut {NoStop}%
\bibitem [{\citenamefont {Farmer}(1982)}]{farmer1982information}%
  \BibitemOpen
  \bibfield  {author} {\bibinfo {author} {\bibfnamefont {J.~D.}\ \bibnamefont
  {Farmer}},\ }\href@noop {} {\bibfield  {journal} {\bibinfo  {journal}
  {Zeitschrift f{\"u}r Naturforschung A}\ }\textbf {\bibinfo {volume} {37}},\
  \bibinfo {pages} {1304} (\bibinfo {year} {1982})}\BibitemShut {NoStop}%
\bibitem [{\citenamefont {Besag}(1974)}]{besag1974spatial}%
  \BibitemOpen
  \bibfield  {author} {\bibinfo {author} {\bibfnamefont {J.}~\bibnamefont
  {Besag}},\ }\href@noop {} {\bibfield  {journal} {\bibinfo  {journal} {Journal
  of the Royal Statistical Society. Series B (Methodological)}\ ,\ \bibinfo
  {pages} {192}} (\bibinfo {year} {1974})}\BibitemShut {NoStop}%
\bibitem [{\citenamefont {Nguyen}\ \emph {et~al.}(2017)\citenamefont {Nguyen},
  \citenamefont {Zecchina},\ and\ \citenamefont {Berg}}]{nguyen2017inverse}%
  \BibitemOpen
  \bibfield  {author} {\bibinfo {author} {\bibfnamefont {H.~C.}\ \bibnamefont
  {Nguyen}}, \bibinfo {author} {\bibfnamefont {R.}~\bibnamefont {Zecchina}}, \
  and\ \bibinfo {author} {\bibfnamefont {J.}~\bibnamefont {Berg}},\ }\href@noop
  {} {\bibfield  {journal} {\bibinfo  {journal} {Advances in Physics}\ }\textbf
  {\bibinfo {volume} {66}},\ \bibinfo {pages} {197} (\bibinfo {year}
  {2017})}\BibitemShut {NoStop}%
\bibitem [{\citenamefont {Crutchfield}\ \emph {et~al.}(2009)\citenamefont
  {Crutchfield}, \citenamefont {Ellison},\ and\ \citenamefont
  {Mahoney}}]{crutchfield2009time}%
  \BibitemOpen
  \bibfield  {author} {\bibinfo {author} {\bibfnamefont {J.~P.}\ \bibnamefont
  {Crutchfield}}, \bibinfo {author} {\bibfnamefont {C.~J.}\ \bibnamefont
  {Ellison}}, \ and\ \bibinfo {author} {\bibfnamefont {J.~R.}\ \bibnamefont
  {Mahoney}},\ }\href@noop {} {\bibfield  {journal} {\bibinfo  {journal}
  {Physical review letters}\ }\textbf {\bibinfo {volume} {103}},\ \bibinfo
  {pages} {094101} (\bibinfo {year} {2009})}\BibitemShut {NoStop}%
\bibitem [{\citenamefont {Ellison}\ \emph {et~al.}(2009)\citenamefont
  {Ellison}, \citenamefont {Mahoney},\ and\ \citenamefont
  {Crutchfield}}]{ellison2009prediction}%
  \BibitemOpen
  \bibfield  {author} {\bibinfo {author} {\bibfnamefont {C.~J.}\ \bibnamefont
  {Ellison}}, \bibinfo {author} {\bibfnamefont {J.~R.}\ \bibnamefont
  {Mahoney}}, \ and\ \bibinfo {author} {\bibfnamefont {J.~P.}\ \bibnamefont
  {Crutchfield}},\ }\href@noop {} {\bibfield  {journal} {\bibinfo  {journal}
  {Journal of Statistical Physics}\ }\textbf {\bibinfo {volume} {136}},\
  \bibinfo {pages} {1005} (\bibinfo {year} {2009})}\BibitemShut {NoStop}%
\bibitem [{\citenamefont {Crutchfield}\ and\ \citenamefont
  {Ellison}(2010)}]{crutchfield2010past}%
  \BibitemOpen
  \bibfield  {author} {\bibinfo {author} {\bibfnamefont {J.~P.}\ \bibnamefont
  {Crutchfield}}\ and\ \bibinfo {author} {\bibfnamefont {C.~J.}\ \bibnamefont
  {Ellison}},\ }\href@noop {} {\bibfield  {journal} {\bibinfo  {journal} {arXiv
  preprint arXiv:1012.0356}\ } (\bibinfo {year} {2010})}\BibitemShut {NoStop}%
\bibitem [{\citenamefont {Feldman}\ and\ \citenamefont
  {Crutchfield}(1998)}]{feldman1998discovering}%
  \BibitemOpen
  \bibfield  {author} {\bibinfo {author} {\bibfnamefont {D.~P.}\ \bibnamefont
  {Feldman}}\ and\ \bibinfo {author} {\bibfnamefont {J.~P.}\ \bibnamefont
  {Crutchfield}}\ }(\bibinfo {organization} {Santa Fe Institute},\ \bibinfo
  {year} {1998})\BibitemShut {NoStop}%
\end{thebibliography}%

\end{document}